\documentclass{article}
\usepackage{amssymb}
\usepackage{amsfonts}
\usepackage{amsmath}

\newtheorem{theorem}{Theorem}
\newtheorem{acknowledgement}[theorem]{Acknowledgement}

\input{tcilatex}

\begin{document}

\title{Validity of Quantum Adiabatic Theorem}
\author{Zhaoyan Wu and Hui Yang \\
%EndAName
Center for Theoretical Physics, Jilin University\\
Changchun, Jilin 130023, China}
\maketitle

\begin{abstract}
The consistency of quantum adiabatic theorem has been doubted recently. It
is shown in the present paper, that the difference between the adiabatic
solution and the exact solution to the Schr\"{o}dinger equation with a
slowly changing driving Hamiltonian is small; while the difference between
their time derivatives is not small. This explains why substituting the
adiabatic solution back into Schr\"{o}dinger equation leads to
`inconsistency' of the adiabatic theorem. Physics is determined completely
by the state vector, and not by its time derivative. Therefore the quantum
adiabatic theorem is physically correct.
\end{abstract}

\subsection{Introduction}

Quantum adiabatic theorem (QAT) dates back to the early years of quantum
mechanics[1]. It has important applications within and beyond quantum
physics. In 1984, M. Berry found there is a geometrical phase in the
adiabatically evolving wavefunction besides the dynamic phase[2]. B. Simon
pointed out Berry's phase factor is the holonomy of a Hermitian line
bundle[3]. This started a rush for geometrical phases in quantum physics[4],
which helped people to get deeper insight into many physical phenomena, such
as Bohm-Aharanov effect, quantum Hall effect, etc. Recently, the quantum
adiabatic theorem has renewed its importance in the context of quantum
control and quantum computation[5]. More recently, however, the consistency
of QAT has been doubted[6]. In their paper entitled "Inconsistency in the
application of the adiabatic theorem", K.-P. Marzlin and B.C. Sanders gave a
proof of inconsistency of QAT, and declared that the standard treatment of
QAT alone does not ensure that a formal application of it result in correct
results. This interesting suggestion has attracted attention from the
physics circle[7]. The purpose of this letter is to point out that the QAT
does give approximate state-vectors\ (wavefunctions), but not necessarily
the approximate time derivatives of state-vectors. While physics is
completely determind by the state-vector, and has nothing to do with its
time derivative. Therefore the QAT is physically correct. What leads to
`inconsistency' of the QAT is neglecting the fact that the adiabatic
approximate state-vector does not necessarily give the approximate time
derivative of state-vector ($\left\Vert \psi _{A}(t)-\psi (t)\right\Vert \ll
1\nRightarrow \left\Vert \overset{\cdot }{\psi }_{A}(t)-\overset{\cdot }{%
\psi }(t)\right\Vert \ll 1,$ where $\left\Vert \varphi \right\Vert \equiv 
\sqrt{\langle \varphi |\varphi \rangle }$ denotes the norm of state-vector $%
\varphi $ ).

\subsection{Standard treatment of QAT}

Suppose that the Hamiltonian depends on $N$ real parameters $R^{1},\ldots
,R^{N}:$%
\begin{equation}
H=H(R^{1},\ldots ,R^{N})=H(R)  \label{eqn1}
\end{equation}%
When the representing point of the Hamiltonian describes slowly a finite
curve $C$ on the $N-$dimenssional parameter manifold $\mathcal{M}$%
\begin{equation}
C:R^{\sigma }=R^{\sigma }(t),\ \forall t\in \lbrack 0,T],1\leq \sigma \leq N
\end{equation}%
where $T$ is the evolution time, let us study the evolution of the system.
The instantaneous Hamiltonian's eigen equation is%
\begin{equation}
H(R)u_{n}(R)=E_{n}(R)u_{n}(R)
\end{equation}%
Getting to the rotating representation%
\begin{equation}
\psi (t)=\sum_{n\geq 0}c_{n}(t)u_{n}(R(t))\exp [\frac{-i}{\hbar }%
\int_{0}^{t}E_{n}(R(t^{\prime }))dt^{\prime }]
\end{equation}%
we get the Schr\"{o}dinger equation

$\ \ \ \ \ \ \overset{\cdot }{c_{m}}(t)=-\sum_{n\geq 0}\left\langle
u_{m}(R(t))\mid \overset{\cdot }{u}_{n}(R(t))\right\rangle \times $%
\begin{equation}
\exp \left\{ \frac{i}{\hbar }\int_{0}^{t}[E_{m}(R(t^{\prime
}))-E_{n}(R(t^{\prime }))]dt^{\prime }\right\} c_{n}(t)
\end{equation}%
To avoid confusion of infinitesimals of different orders and to show what
`rapidly oscillating' means, let's change to the dimensionless time $\tau
=t/T$

\QTP{Body Math}
$\ \ \ \ \ \frac{d}{d\tau }\check{c}_{m}(\tau )=-\sum_{n\geq 0}\left\langle
u_{m}(\check{R}(\tau ))\mid \frac{d}{d\tau }u_{n}(\check{R}(\tau
))\right\rangle \times $%
\begin{equation}
\exp \left\{ \frac{i}{\hbar }T\int_{0}^{\tau }[E_{m}(\check{R}(\tau ^{\prime
}))-E_{n}(\check{R}(\tau ^{\prime }))]d\tau ^{\prime }\right\} \check{c}%
_{n}(\tau )
\end{equation}%
where%
\begin{equation}
\check{c}_{m}(\tau )=c_{m}(T\tau )=c_{m}(t),\check{R}(\tau )=R(T\tau )=R(t)
\end{equation}%
The initial value problem of the above differential equations is equivalent
to the following integral equations

\QTP{Body Math}
$\ \ \ \ \ \ \check{c}_{m}(\tau )=\check{c}_{m}(0)-\sum_{n\geq
0}\int_{0}^{\tau }\left\langle u_{m}(\check{R}(\tau _{1}))\mid \frac{d}{%
d\tau }u_{n}(\check{R}(\tau _{1}))\right\rangle \times $%
\begin{equation}
\exp \left\{ \frac{i}{\hbar }T\int_{0}^{\tau _{1}}[E_{m}(\check{R}(\tau
^{\prime }))-E_{n}(\check{R}(\tau ^{\prime }))]d\tau ^{\prime }\right\} 
\check{c}_{n}(\tau _{1})d\tau _{1}
\end{equation}%
Let's slow down evenly the changing speed of the Hamiltonian while keep the
finite curve $C$ fixed. Mathematically, that is to\ let $T\rightarrow \infty 
$, while keeping the function form of $\check{R}(\tau )$ unchanged. The
oscillating factors in the integrand ensure vanishing of the corresponding
integrals. There is no resonance problem in the mathematical context. For
the practical physical problem, slowly changing of the Hamiltonian means $T$
is such a long time that%
\begin{equation}
\left\vert \frac{\left\langle u_{m}(R(t))\mid \overset{\cdot }{u}%
_{n}(R(t))\right\rangle \hbar }{E_{m}(R(t))-E_{n}(R(t))}\right\vert \ll 1
\end{equation}%
In both mathematics and physics contexts, the integral equation (8) can be
approximately rewritten as%
\begin{equation}
\check{c}_{m}^{A}(\tau )=\check{c}_{m}(0)-\int_{0}^{\tau }\left\langle u_{m}(%
\check{R}(\tau _{1}))\mid \frac{d}{d\tau }u_{m}(\check{R}(\tau
_{1}))\right\rangle \check{c}_{m}^{A}(\tau _{1})d\tau _{1}
\end{equation}%
Solving this equation by using iteration gives%
\begin{equation}
\check{c}_{m}^{A}(\tau )=\exp \left\{ -\int_{0}^{\tau }\left\langle u_{m}(%
\check{R}(\tau _{1}))\mid \frac{d}{d\tau }u_{m}(\check{R}(\tau
_{1}))\right\rangle d\tau _{1}\right\} \check{c}_{m}(0)
\end{equation}%
This proves the QAT.

\subsection{Analysis of `Inconsistency' of QAT}

When we substitute the adiabatic approximate solution (11) back into the
integral equations (8), the equations approximately hold.

\QTP{Body Math}
$\ \ \ \ \ 0\approx -\sum_{n(\neq m)}\int_{0}^{\tau }\left\langle u_{m}(%
\check{R}(\tau _{1}))\mid \frac{d}{d\tau }u_{n}(\check{R}(\tau
_{1}))\right\rangle \times $%
\begin{equation}
\exp \left\{ \frac{i}{\hbar }T\int_{0}^{\tau _{1}}[E_{m}(\check{R}(\tau
^{\prime }))-E_{n}(\check{R}(\tau ^{\prime }))]d\tau ^{\prime }\right\} 
\check{c}_{n}^{A}(\tau _{1})d\tau _{1}
\end{equation}%
However, when we substitute the adiabatic approximate solution (11) back
into the differential equations (6) whose initial value problem is
equivalent to the integral equations (8), we obtain

$\ \ \ \ \ \ \ 0\approx -\sum_{n(\neq m)}\left\langle u_{m}(\check{R}(\tau
))\mid \frac{d}{d\tau }u_{n}(\check{R}(\tau ))\right\rangle \times $%
\begin{equation}
\exp \left\{ \frac{i}{\hbar }T\int_{0}^{\tau }[E_{m}(\check{R}(\tau ^{\prime
}))-E_{n}(\check{R}(\tau ^{\prime }))]d\tau ^{\prime }\right\} \check{c}%
_{n}^{A}(\tau )
\end{equation}%
Considering that $\psi (0)$ can be an arbitrary state-vector, we have

\begin{equation}
0\approx \left\langle u_{m}(\check{R}(\tau ))\mid \frac{d}{d\tau }u_{n}(%
\check{R}(\tau ))\right\rangle ,\forall m\neq n
\end{equation}%
which is false. Notice that the right-hand side of (13) is the derivative of
the right-hand side of (12), while (12) is correct and (13) is incorrect. In
order to understand the situation we are facing, let's study the following
basic mathematical fact. Let $|\psi (t)\rangle \equiv |0\rangle e^{-i\omega
t}+\varepsilon |1\rangle e^{-it/(\varepsilon ^{2})},(0<\varepsilon \ll 1),$ $%
|\varphi (t)\rangle \equiv |0\rangle e^{-i\omega t},$ where $|0\rangle
,|1\rangle $ are eigen vectors of the 1-dimensional harmonic oscillator
energy.%
\begin{equation}
\because \left\Vert |\psi (t)\rangle -|\varphi (t)\rangle \right\Vert
=\varepsilon \ll 1,\therefore |\psi (t)\rangle \approx |\varphi (t)\rangle
\end{equation}%
While 
\begin{equation}
\because \left\Vert |\overset{\cdot }{\psi }(t)\rangle -|\overset{\cdot }{%
\varphi }(t)\rangle \right\Vert =1/\varepsilon \gg 1,\therefore |\overset{%
\cdot }{\psi }(t)\rangle \nsim |\overset{\cdot }{\varphi }(t)\rangle
\end{equation}%
The above example shows that two approximately equal time-dependent
state-vectors do not necessarily have approximately equal time derivatives.
Therefore the approximate solution to integral equations (8) does not ensure
that the equivalent differential equations (6) approximately hold. It is
neglect of this basic mathematical fact that leads to `Inconsistency' of QAT
in [6].

QAT gives the approximate state-vector, not the approximate time derivative
of the state-vector. All the physics is, however, determined by the
state-vector itself, not by its time derivative. Therefore QAT is completely
correct physically.

\subsection{An exactly solvable example}

\QTP{Body Math}
Let's consider an exactly solvable example, the evolution of the spin
wavefunction of an electron in a slowly rotating magnetic field$%
\overrightarrow{B}(t)=B_{0}(\overrightarrow{i}\cos \frac{2\pi }{T}t+%
\overrightarrow{j}\sin \frac{2\pi }{T}t)$.The instantaneous Hamiltonian is

\QTP{Body Math}
$\ \ \ \ \ \ \ \ \ \ \ \ \ H(t)=-\overrightarrow{\mu }\cdot \overrightarrow{B%
}(t)=\frac{e}{m}\overrightarrow{s}\cdot \overrightarrow{B}(t)=\frac{e\hbar }{%
2m}\overrightarrow{\sigma }\cdot \overrightarrow{B}(t)$%
\begin{equation}
=\frac{e\hbar B_{0}}{2m}\left[ 
\begin{array}{cc}
0 & e^{^{^{-i2\pi t/T}}} \\ 
e^{^{i2\pi t/T}} & 0%
\end{array}%
\right] =\varepsilon \left[ 
\begin{array}{cc}
0 & e^{^{-i2\pi t/T}} \\ 
e^{^{i2\pi t/T}} & 0%
\end{array}%
\right]
\end{equation}%
Its eigenvalues are $E_{\pm }(t)=\pm \varepsilon $. And the corresponding
eigenvectors are

\QTP{Body Math}
\begin{equation}
u_{\pm }(t)=\frac{1}{\sqrt{2}}\left[ 
\begin{array}{c}
e^{-i\pi t/T} \\ 
\pm e^{i\pi t/T}%
\end{array}%
\right]
\end{equation}%
The exact general solution to the Schrodinger equation

\QTP{Body Math}
\begin{equation}
i\hbar \frac{d}{dt}\psi (t)=H(t)\psi (t)
\end{equation}%
or

\begin{equation}
i\hbar \frac{d}{dt}\left[ 
\begin{array}{c}
x(t) \\ 
y(t)%
\end{array}%
\right] =\varepsilon \left[ 
\begin{array}{cc}
0 & e^{^{-i2\pi t/T}} \\ 
e^{i2\pi t/T} & 0%
\end{array}%
\right] \left[ 
\begin{array}{c}
x(t) \\ 
y(t)%
\end{array}%
\right]
\end{equation}%
is

$\ \ \ \ \ \ \ \ \ \ \psi (t)=\left[ 
\begin{array}{c}
x(t) \\ 
y(t)%
\end{array}%
\right] $%
\begin{equation}
=\left[ 
\begin{array}{c}
-A^{-1}[c_{1}(B+C)e^{iCt}+c_{2}(B-C)e^{-iCt}]e^{-iBt} \\ 
\lbrack c_{1}e^{iCt}+c_{2}e^{-iCt}]e^{^{iBt}}%
\end{array}%
\right]
\end{equation}%
where $A=\varepsilon /\hbar ,B=\pi /T,C=\sqrt{A^{2}+B^{2}}$, and $c_{1}$, $%
c_{2}$ are the integral constants. The specific solution determined by the
initial condition%
\begin{equation}
\psi (0)=\left[ 
\begin{array}{c}
x(0) \\ 
y(0)%
\end{array}%
\right] =\frac{1}{\sqrt{2}}\left[ 
\begin{array}{c}
1 \\ 
1%
\end{array}%
\right]
\end{equation}%
is

\QTP{Body Math}
\begin{equation}
\psi (t)=\frac{1}{\sqrt{2}}\left[ 
\begin{array}{c}
\left( \cos Ct-i\frac{A-B}{C}\sin Ct\right) e^{-iBt} \\ 
\left( \cos Ct-i\frac{A+B}{C}\sin Ct\right) e^{^{iBt}}%
\end{array}%
\right]
\end{equation}%
Let's get into the rotating representation.

\QTP{Body Math}
\begin{equation}
\psi (t)=c_{+}(t)u_{+}(t)e^{-iAt}+c_{-}(t)u_{-}(t)e^{iAt}
\end{equation}%
The exact Schr\"{o}dinger equation becomes

\QTP{Body Math}
$\ \ \ \ \ \ \ \ \ \ \ \ \ \ \ \ \ \ \ \ \ \ \ \ \ \ \ \ \ \ \ \ \ \overset{%
\cdot }{c}_{+}(t)=iBe^{i2At}c_{-}(t)$%
\begin{equation}
\overset{\cdot }{c}_{-}(t)=iBe^{-i2At}c_{+}(t)
\end{equation}%
Its general solution is%
\begin{equation}
\left[ 
\begin{array}{c}
c_{+}(t) \\ 
c_{-}(t)%
\end{array}%
\right] =\left[ 
\begin{array}{c}
e^{iAt}\left( \frac{C-A}{B}c^{\prime }e^{iCt}-\frac{C+A}{B}c^{\prime \prime
}e^{-iCt}\right) \\ 
e^{-iAt}\left( c^{\prime }e^{iCt}+c^{\prime \prime }e^{-iCt}\right)%
\end{array}%
\right]
\end{equation}%
The specific solution determined by the initial condition%
\begin{equation}
\left[ 
\begin{array}{c}
c_{+}(0) \\ 
c_{-}(0)%
\end{array}%
\right] =\left[ 
\begin{array}{c}
1 \\ 
0%
\end{array}%
\right]
\end{equation}%
is%
\begin{equation}
\left[ 
\begin{array}{c}
c_{+}(t) \\ 
c_{-}(t)%
\end{array}%
\right] =\left[ 
\begin{array}{c}
\left( \cos Ct-i\frac{A}{C}\sin Ct\right) e^{iAt} \\ 
(i\frac{B}{C}\sin Ct)e^{-iAt}%
\end{array}%
\right]
\end{equation}%
The adiabatic approximation means neglecting the non-diagonal ($n\neq m$)
terms, which contain oscillating factors, on the right-hand side of
differential equations (25). The adiabatic approximate solution determind by
the initial condition (27) is%
\begin{equation}
\left[ 
\begin{array}{c}
c_{+}^{A}(t) \\ 
c_{-}^{A}(t)%
\end{array}%
\right] =\left[ 
\begin{array}{c}
1 \\ 
0%
\end{array}%
\right]
\end{equation}%
Getting to the dimenssionless time $\tau =t/T$, we rewrite (28) and (29) as

\QTP{Body Math}
$\ \ \left[ 
\begin{array}{c}
\check{c}_{+}(\tau ) \\ 
\check{c}_{-}(\tau )%
\end{array}%
\right] =$%
\begin{equation}
\left[ 
\begin{array}{c}
\left( \cos \sqrt{\left( \varepsilon T/\hbar \right) ^{2}+\pi ^{2}}\tau -i%
\frac{\varepsilon T/\hbar }{\sqrt{\left( \varepsilon T/\hbar \right)
^{2}+\pi ^{2}}}\sin \sqrt{\left( \varepsilon T/\hbar \right) ^{2}+\pi ^{2}}%
\tau \right) e^{i\varepsilon T\tau /\hbar } \\ 
\left( i\frac{\pi }{\sqrt{(\varepsilon T/\hbar )^{2}+\pi ^{2}}}\sin \sqrt{%
\left( \varepsilon T/\hbar \right) ^{2}+\pi ^{2}}\tau \right)
e^{-i\varepsilon T\tau /\hbar }%
\end{array}%
\right]
\end{equation}%
\begin{equation}
\left[ 
\begin{array}{c}
\check{c}_{+}^{A}(\tau ) \\ 
\check{c}_{-}^{A}(\tau )%
\end{array}%
\right] =\left[ 
\begin{array}{c}
1 \\ 
0%
\end{array}%
\right]
\end{equation}%
It's easy to see that%
\begin{equation}
\left[ 
\begin{array}{c}
\check{c}_{+}(\tau ) \\ 
\check{c}_{-}(\tau )%
\end{array}%
\right] \underset{T\rightarrow \infty }{\longrightarrow }\left[ 
\begin{array}{c}
1 \\ 
0%
\end{array}%
\right] =\left[ 
\begin{array}{c}
\check{c}_{+}^{A}(\tau ) \\ 
\check{c}_{-}^{A}(\tau )%
\end{array}%
\right]
\end{equation}%
The difference between (30) and (31) is small, but rapidly oscillates with
the dimensionless time $\tau $. Therefore, it's to be expected that the
derivative with $\tau $ of the difference is no longer small. In fact,
letting $F\equiv \sqrt{\left( \varepsilon T/\hbar \right) ^{2}+\pi ^{2}}$,
we have

\QTP{Body Math}
$\left[ 
\begin{array}{c}
\frac{d}{d\tau }\check{c}_{+}(\tau ) \\ 
\frac{d}{d\tau }\check{c}_{-}(\tau )%
\end{array}%
\right] =\left[ 
\begin{array}{c}
\left( \frac{-\pi ^{2}}{F}\sin F\tau \right) e^{i\varepsilon T\tau /\hbar }
\\ 
\left( \frac{\pi \varepsilon T/\hbar }{F}\sin F\tau +i\pi \cos F\tau \right)
e^{-i\varepsilon T\tau /\hbar }%
\end{array}%
\right] $%
\begin{equation}
\underset{T\rightarrow \infty }{\rightarrow }\left[ 
\begin{array}{c}
0 \\ 
i\pi e^{-i2\varepsilon T\tau /\hbar }%
\end{array}%
\right] \neq \left[ 
\begin{array}{c}
0 \\ 
0%
\end{array}%
\right] =\left[ 
\begin{array}{c}
\frac{d}{d\tau }\check{c}_{+}^{A}(\tau ) \\ 
\frac{d}{d\tau }\check{c}_{-}^{A}(\tau )%
\end{array}%
\right]
\end{equation}

\subsection{conclusion}

The above discussion shows: (i) The adiabatic state-vector $\psi ^{A}(t)$
does not satisfies approximately the Schr\"{o}dinger differential equation,
but it satisfies approximately the equivalent integral equation. (ii) The
QAT is completely correct physically. This is ensured by $\left\Vert \psi
(t)-\psi ^{A}(t)\right\Vert \ll 1$. But it's not necessarily true that $%
\left\Vert \overset{\cdot }{\psi }(t)-\overset{\cdot }{\psi ^{A}}%
(t)\right\Vert \ll 1$. (iii) Taking $\overset{\cdot }{\psi ^{A}}(t)$ for $%
\overset{\cdot }{\psi }(t)$ will probably lead to contradiction.

Even though we don't agree with [6], we still think it's an interesting
work. Because it has raised an important question: In theoretical reasoning,
one has to bear in mind that approximately equal functions do not have to
have approximately equal derivatives.

\begin{acknowledgement}
We thank L.-A. Wu for letting us know the work [6] and helpful discussion.
\end{acknowledgement}

\end{document}